\documentclass[a4paper,10pt,twoside]{cpc-hepnp}

\usepackage{multicol}
\usepackage{graphicx}
\usepackage{booktabs}
\usepackage{amssymb,bm,mathrsfs,bbm,amscd}
\usepackage[tbtags]{amsmath}
\usepackage{lastpage}
\usepackage{CJK}
\usepackage{multicol,enumerate}
\usepackage{multirow,amsmath,amssymb}
\usepackage{subfigure}
\usepackage{slashed}
\usepackage{color}

\newcommand{\tabincell}[2]{\begin{tabular}{@{}#1@{}}#2\end{tabular}}
\newcommand{\whizard}{{\sc Whizard}}
\newcommand{\pythia}{{\sc Pythia}}
\newcommand{\fbinv} {\mbox{\ensuremath{~\text{fb}^\text{$-$1}}}}
\newcommand{\higgs} {\mathrm{H}}
\def\mumu{\ensuremath{\mathrm{\mu^+ \mu^-}}}%

\begin{document}
\begin{CJK*}{GBK}{song}

\fancyhead[c]{\small Prepared for submission to Chinese Physics C~~~
} \fancyfoot[C]{\small -\thepage}

\footnotetext[0]{Received 28 November 2017}

\title{Measurement of $\higgs \xrightarrow {}\mumu$ production in association with a Z boson at the CEPC\thanks{Supported by National Natural Science Foundation of China (11475190) and National Natural Science Foundation of China (11575005) }}

\author{%
      Zhen-Wei Cui$^{1;1)}$\email{lovekey@pku.edu.cn}%
\quad Qiang Li$^{1,2;2)}$\email{qliphy0@pku.edu.cn}%
\quad Gang Li$^{2,3}$
\quad Man-Qi Ruan$^{2,3}$ 
\quad Lei Wang$^{1}$
\quad Da-Neng Yang$^{1}$
}
\maketitle

\address{%
$^1$ Department of Physics and State Key Laboratory of Nuclear Physics and Technology, Peking University, Beijing, 100871, China\\
$^2$ CAS Center for Excellence in Particle Physics, Beijing 100049, China\\
$^3$ Institute of High Energy Physics, Beijing 100049, China\\
}

\begin{abstract}
The Circular Electron-Positron Collider (CEPC) is a future Higgs factory proposed by the Chinese high energy physics community. It will operate at a center-of-mass energy of 240-250 GeV and is expected to accumulate an integrated luminosity of 5 ab$^{\rm{-1}}$ with ten years of operation. At CEPC, Higgs bosons are dominantly produced from $ZH$ associated process. Vast amount of Higgs events collected will enable precise studies of its properties including Yukawa couplings to massive particles. With {\sc GEANT4}-based simulation of detector effects, we study CEPC feasibility on measuring Higgs boson decaying into a pair of muons.  The results with or without information from Z boson decay products are provided, which show a signal significance of over 10 standard deviations can be achieved and the H-$\mathrm{\mu}$-$\mathrm{\mu}$ coupling can be measured within $10\%$ accuracy.
\end{abstract}



\footnotetext[0]{\hspace*{-3mm}\raisebox{0.3ex}{$\scriptstyle\copyright$}2013
Chinese Physical Society and the Institute of High Energy Physics
of the Chinese Academy of Sciences and the Institute
of Modern Physics of the Chinese Academy of Sciences and IOP Publishing Ltd}%

\begin{multicols}{2}

\section{Introduction}
\qquad The discovery of the Higgs-like boson completes the particle table of the Standard Model (SM) of particle physics.
Up-to-date LHC measurements all indicate that the Higgs boson is indeed highly SM like~\cite{cmshig,lhcsub1,lhcsub2,lhcsub3,lhcsub4,lhcsub5}.
In the SM, Higgs couplings to massive particles are proportional to their mass (square).
Hence the event rate with Higgs couplings to the first and second generation of massive fermions can be very small, making them difficult to measure at the LHC.
The Circular Electron-Positron Collider (CEPC)~\cite{ref:3}, however, is designed to run around $240\sim250$ GeV with an instantaneous luminosity of 2 $\times$ $\rm{10}^{\rm{34}}$ $\rm{cm}^{\rm{-2}}$ $\rm{s}^{\rm{-1}}$, and will deliver 5 $\mathrm{ab}^{-1}$ of integrated luminosity with ten years of running.
The huge amount of data will enable precise measurement of the Higgs to light fermions branching ratios and determine associated Yukawa couplings, including H-$\mathrm{\mu}$-$\mathrm{\mu}$, which is crucial to validate consistency of the SM Higgs mechanism since any deviation indicates the existence of new physics.

Searches for the $\higgs \to \mathrm{\mu^+ \mu^-}$ production have been performed at ATLAS and CMS experiments with Run-I and Run-II data~\cite{Aad:2014xva,Khachatryan:2014aep,Aaboud:2017ojs}. The most stringent observed (expected) upper limit on the cross-section times branching ratio is found to be 2.8 (2.9) times the SM prediction~\cite{Aaboud:2017ojs}.
Projections have also been made at High Luminosity-LHC assuming an integrated luminosity of 3000 fb$^{\rm{-1}}$ collected by the ATLAS or CMS detector,
 which can lead to a signal significance of about 7 $\sigma$~\cite{ATLprojection} with the H-$\mathrm{\mu}$-$\mathrm{\mu}$ coupling determined with an accuracy of around 20\%~\cite{CMS:2013xfa}.
Studies have also been performed for the International Linear Collider (ILC).
Considering a center mass energy of 250 GeV and an integrated luminosity of 250 fb$^{\rm{-1}}$,
the signal is dominated by the Higgs-strahlung from a Z boson
and the signal significances for the sub-processes with Z boson decays into $\nu\bar{\nu}$ and q$\bar{\mathrm{q}}$ are found to be 1.8 and 1.1 $\sigma$, respectively~\cite{Aihara:2009ad}.
Assuming polarized beams and collisions at a center mass energy 1 TeV with an integrated luminosity of 500 fb$^{\rm{-1}}$,
the signal is dominated by the WW-fusion process and a sensitivity of 2.75 $\sigma$ can be achieved~\cite{Giannelli:2016rja}.

At the CEPC, the signal $\higgs \to \mathrm{\mu^+ \mu^-}$ production is dominated by the Higgs-strahlung from a Z boson.
We perform a feasibility study based on events generated at leading order accuracy with initial state radiation (ISR), parton shower, hadronization and detector effects simulated.

Considering that 70\% of the Z bosons decay hadronically and 20\% decay invisibly,
we focous on two scenarios, one for Z boson inclusive decay and the other for hadronic decay.
The first case maximally exploited the statistics of the produced $\higgs \to \mathrm{\mu^+ \mu^-}$ events and the second category takes advantage of the major part of the decay kinematics. For both cases, we first perform a cut-based analysis and then improve the measurement using a Boosted Decision Tree (BDT) technique.

This paper is organized as follows. Section 2 describes event generation and simulation. Section 3 presents results for the inclusive measurement.
Section 4 presents results for the Z$\to$q$\bar{\mathrm{q}}$ decay channel. Section 5 summarizes the paper.

\section{Monte Carlo Simulation}
\label{mcsim}

\qquad At 250~GeV CEPC, Higgs bosons are mainly produced through Higgs-strahlung,
i.e. $e^+e^- \rightarrow \mathrm{Z} \higgs$.
With an integrated luminosity of 5000 fb$^{\rm{-1}}$,
about 230 of our signal events $\higgs \rightarrow \mu^+\mu^-$ can be produced.
The expected background to the signal production includes
2-fermion processes $e^+e^-\rightarrow f\bar{f}$,
where $f$ can be any SM fermion other than the top quark,
and 4-fermion processes, which can be mediated through associated ZZ, WW, ZZ, WW production and a single Z boson production.
All Monte Carlo (MC) events are generated with \whizard~ V1.9.5~\cite{ref:4} event generator at parton level with ISR and interference effects included.
The generated events are interfaced to \pythia\ 6~\cite{pythia} for parton shower and hadronization simulation.
Detector effects are simulated with the CEPC detector implemented with Mokka/GEANT4~\cite{Mokka,ref:3,Chen:2016zpw}.
The detector is assumed to have a similar structure as the International Large Detector (ILD)~\cite{ild1,ild2} at the ILC~\cite{ILC}.
At the CEPC, the muon identification efficiency is expected to be over 99.5\% for ${\rm{P}}_{\rm{T}}$ larger than 10 GeV, and with excellent ${\rm{P}}_{\rm{T}}$ resolution of $\sigma_{1/{\rm{P}}_{\rm{T}}} = 2\times 10^{-5} \oplus 1\times 10^{-3}/({\rm{P}}_{\rm{T}}\sin\theta)$.
The fully simulated events are reconstructed with a particle-flow algorithm ArborPFA~\cite{arbor}.
More details about the CEPC sample set can be found in reference~\cite{Mo:2015mza}.

The major SM backgrounds, including all the 2-fermion processes($e^+e^-\rightarrow f\bar{f}$,
where $f\bar{f}$ refers to all lepton and quark pairs except $t\bar{t}$) and 4-fermion processes($ZZ$, $WW$, $ZZ$ or $WW$, single $Z$). The initial states radiation (ISR) and all possible interference effects are taken into account in the generation automatically. The classification for four fermions production, is referred to LEP~\cite{ref:17}, depending crucially on the final state.
For example, if the final states consist of two mutually charge conjugated fermion pairs that could decay from both $WW$ and $ZZ$ intermediate state, such as $e^{+}e^{-}\nu_{e}\bar{\nu_{e}}$,
this process is classified as ``$ZZ$ or $WW$'' process.
If there are $e^{\pm}$ together with its parter neutrino and an on-shell $W$ boson in the final tate, this type is named as ``single $W$''.
Meanwhile, if there are a electron-positron pair and a on-shell $Z$ boson  in the final state, this case is named as ``single $Z$''.  Detailed information on the 2-fermion and 4-fermion samples used in our analyses are listed in Tables~\ref{tab:example1} and \ref{tab:example2}.

\section{Inclusive analysis}
\label{inclusive}

\qquad A recoil mass method enables a measurement of the $H\rightarrow \mu^+\mu^-$ production without measuring the associated Z boson decay.
We define the recoil mass as

\begin{equation}
\mathrm{M}_{\text{recoil}}^{2}=s+\mathrm{M}^{2}_{\mathrm{H}}-2\cdot E_{\mathrm{H}}\cdot~ \sqrt[]s ~,
\end{equation}

where $\sqrt[]s$ is the center of mass energy, $\mathrm{M_{H}}$ and $E_{\mathrm{H}}$ correspond to the reconstructed mass and energy of the Higgs boson.
The ZH ($\higgs \to \mu^+\mu^-$) events form a peak in the $M_{\text{recoil}}$ distribution at the Z boson mass window.

We select two muons with largest transverse momenta and
consider selections on the following kinematic variables:
invariant mass of the di-muon system $\mathrm{M}_{\mu^+ \mu^-}$,
recoil mass of the di-moun system $\mathrm{M}_{\text{recoil}}^{\mu^+\mu^-}$,
transverse momentum of the di-muon system $\mathrm{P}_{\mathrm{T}_{\mu^+ \mu^-}}$,
 third component of the di-muon momentum $\mathrm{P}_{\mathrm{Z}_{\mu^+ \mu^-}}$,
energy of di-muon system $\mathrm{E}_{\mu^+ \mu^-}$, and angular variables $\cos \theta_{\mu^-}$, $\cos \theta_{\mu^+}$,
$\cos\theta_{\mu^+ \mu^-}$,
$\cos\theta_{\mathrm{Z}\mu^-}$, and
$\cos\theta_{\mathrm{Z}\mu^+}$, where $\theta_{\mathrm{Z}\mu^{\pm}}$ represents angle between Z boson and muon leptons.

\begin{center}

\begin{minipage}[c]{0.5\textwidth}
 \includegraphics[angle=0,width=4cm]{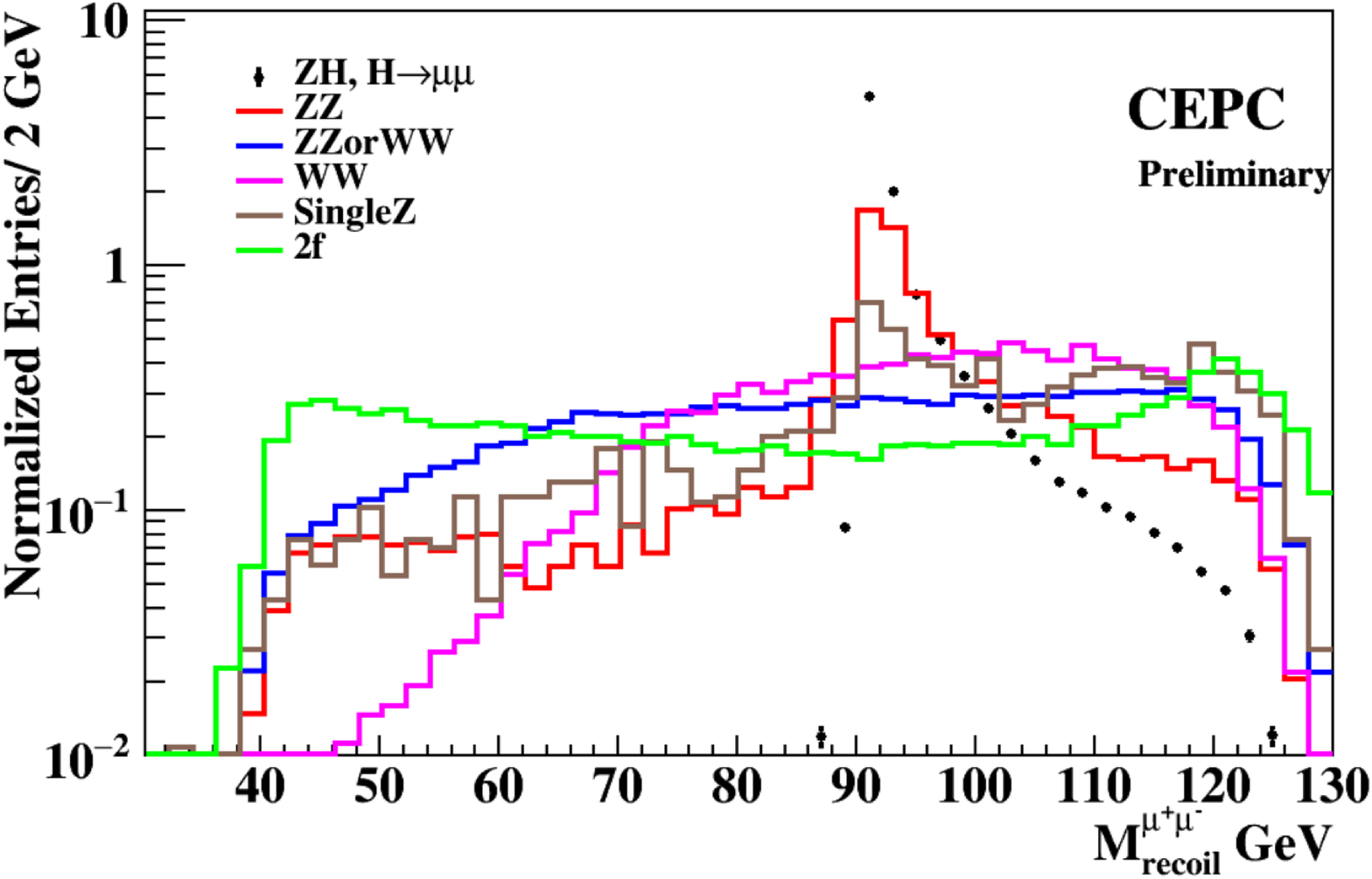}
 \includegraphics[angle=0,width=4cm]{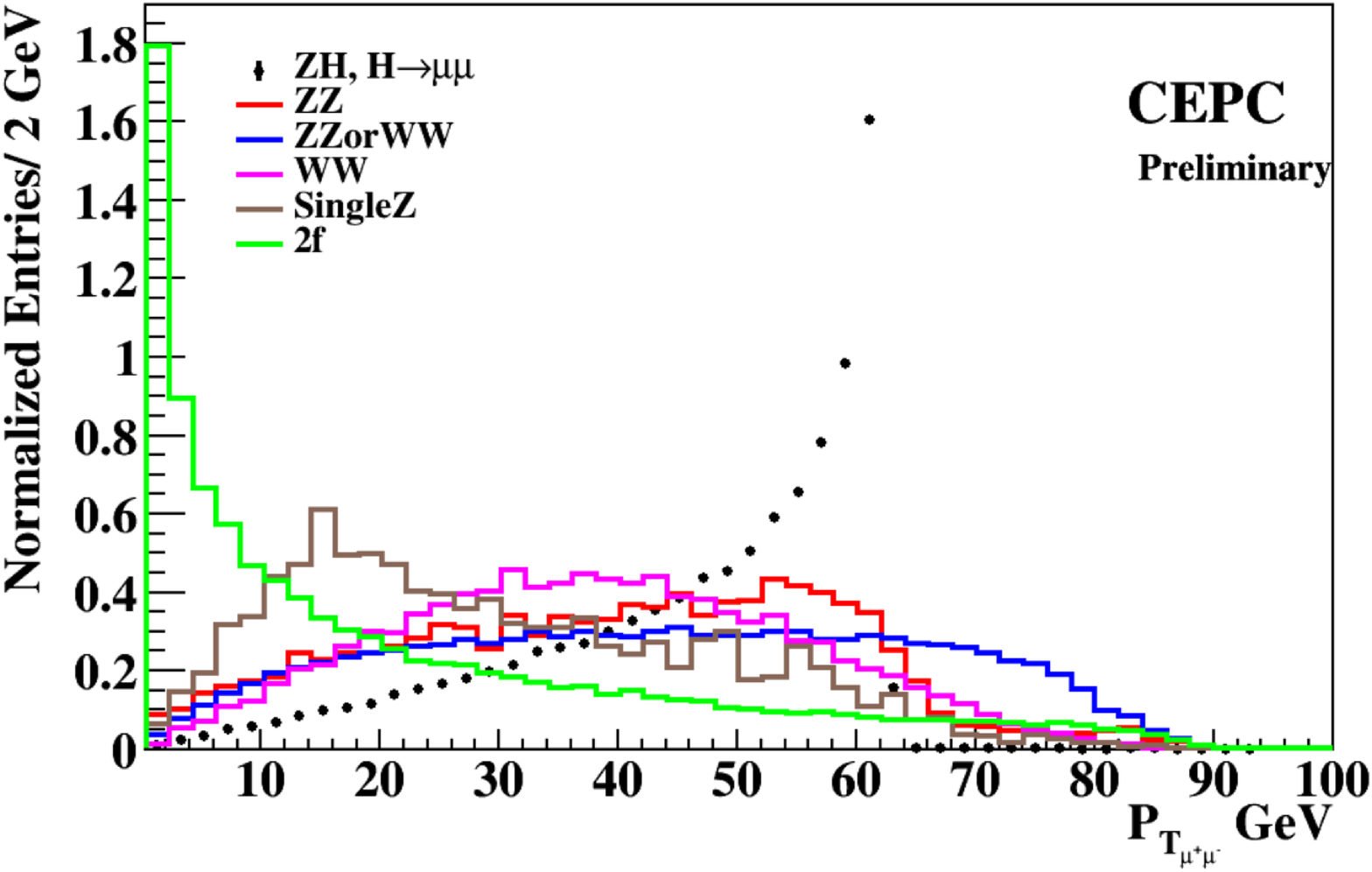}
\label{fig:levfig}
\end{minipage}

\begin{minipage}[c]{0.5\textwidth}
\includegraphics[angle=0,width=4cm]{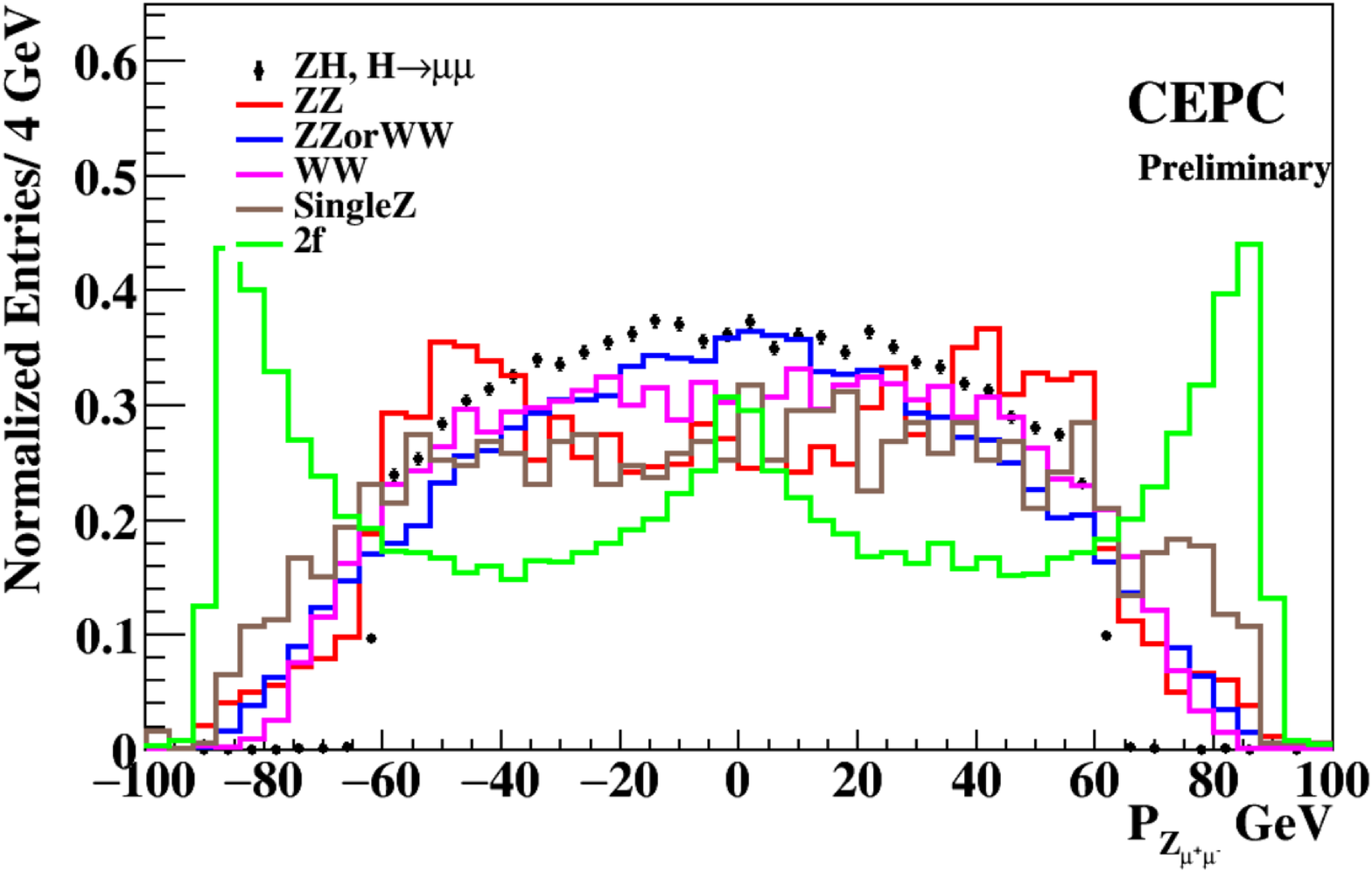}
 \includegraphics[angle=0,width=4cm]{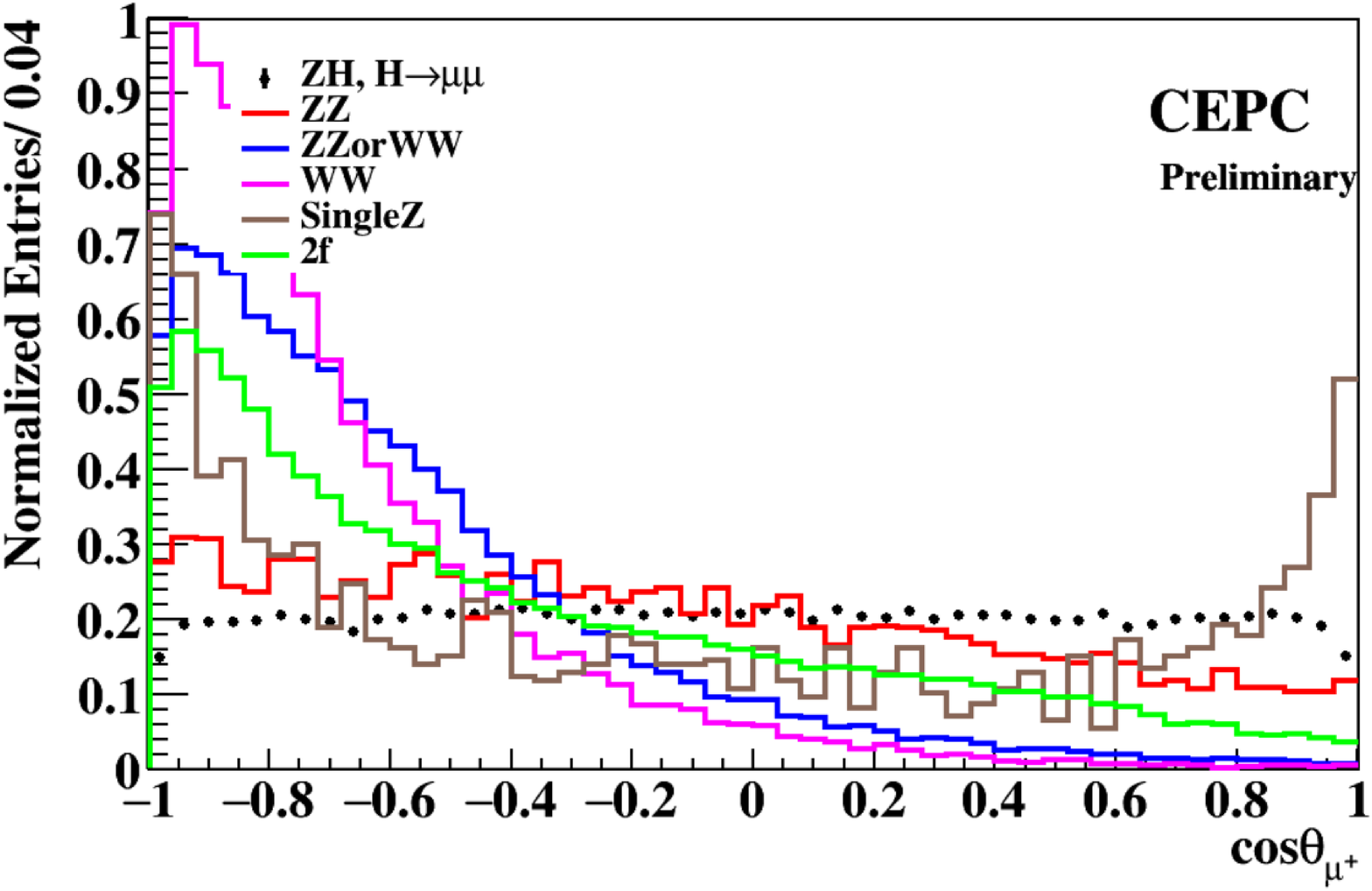}
\label{fig:levfig}
\end{minipage}%

\begin{minipage}[c]{0.5\textwidth}
\includegraphics[angle=0,width=4cm]{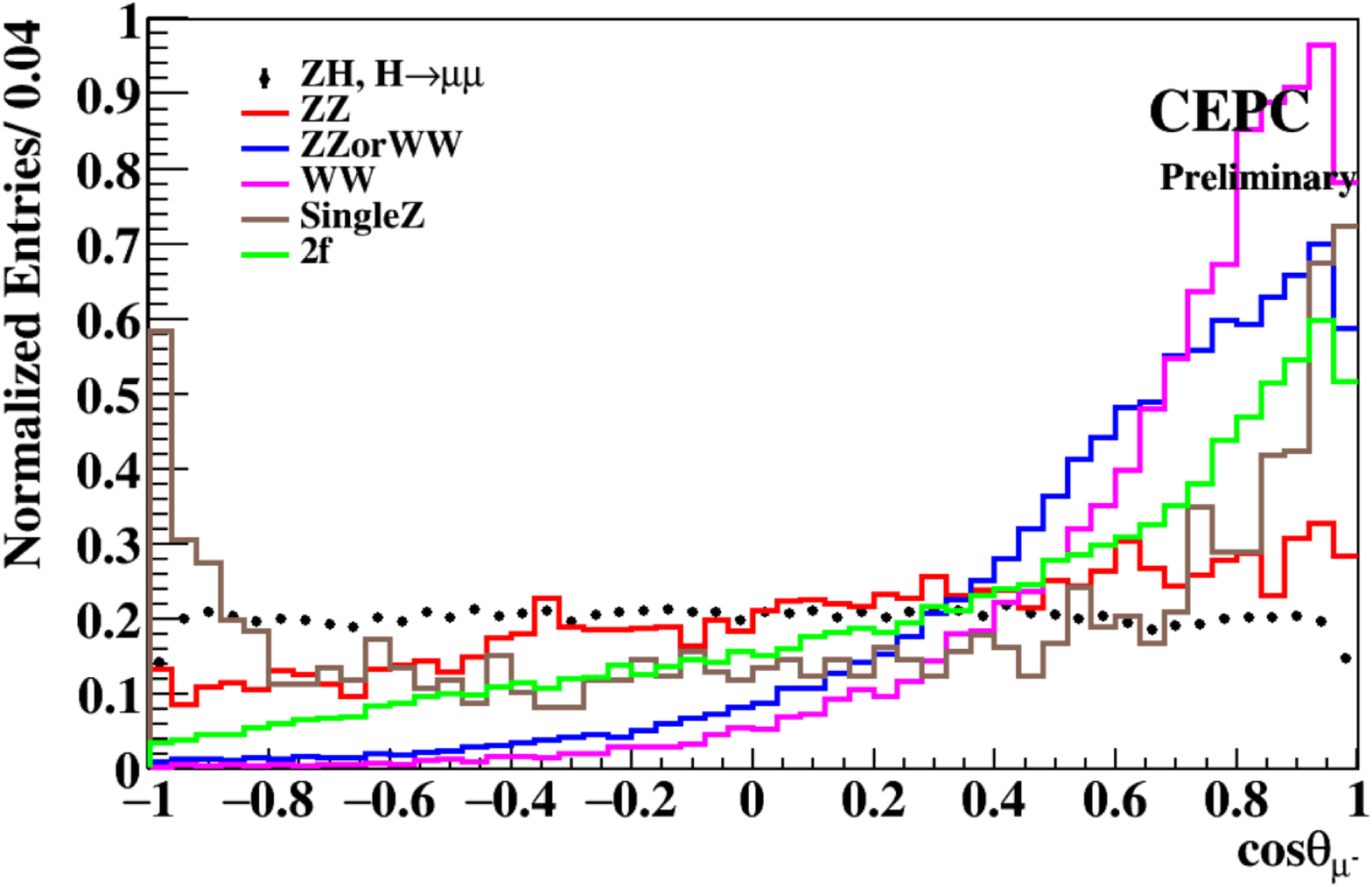}
\includegraphics[angle=0,width=4cm]{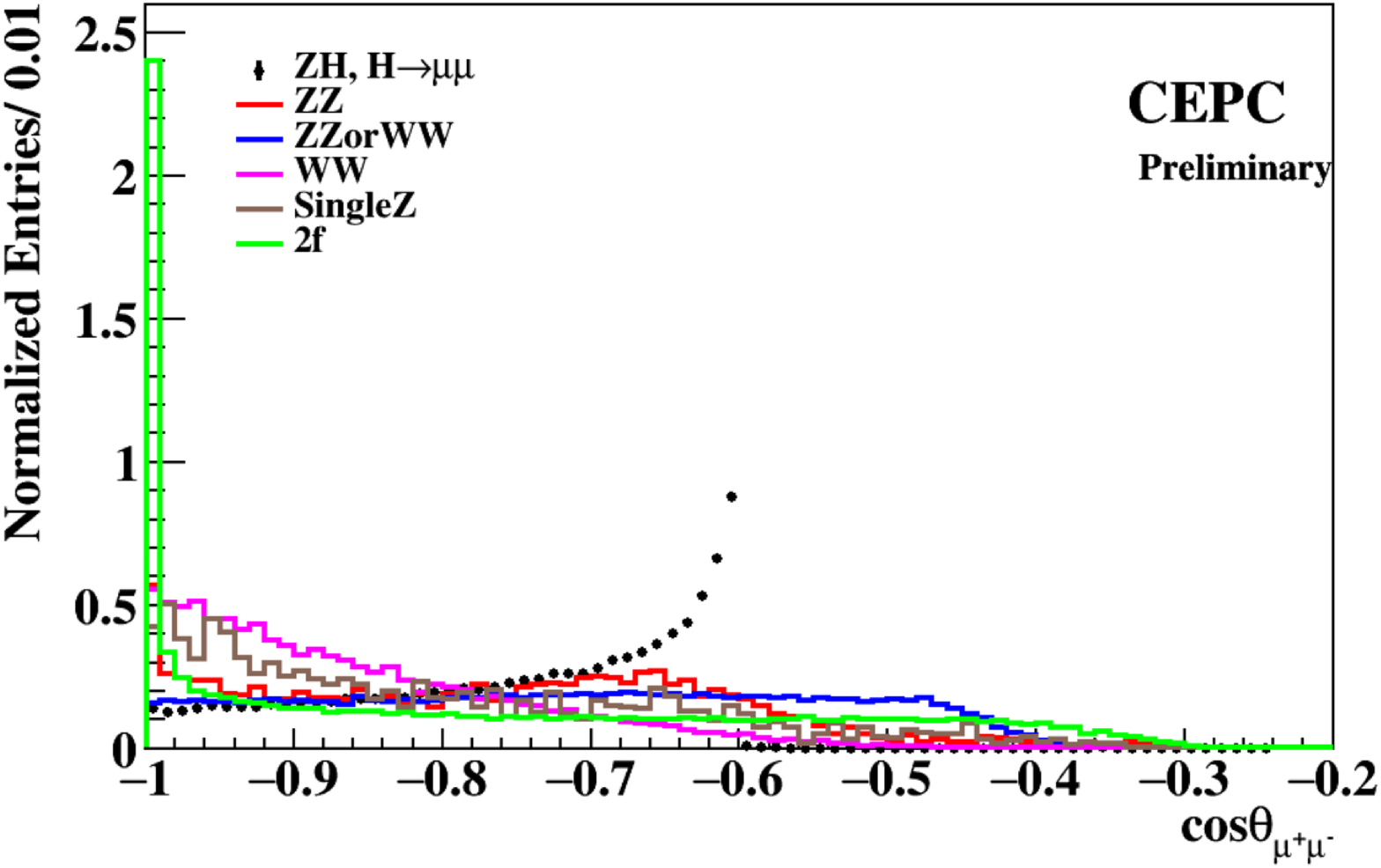}
\label{fig:levfig}
\end{minipage}

\figcaption{Distribution of the $\mathrm{M}_{\text{recoil}}^{\mu^+\mu^-}$, $\mathrm{P}_{\mathrm{T}_{\mu^+ \mu^-}}$, $\mathrm{P}_{\mathrm{Z}_{\mu^+ \mu^-}}$, $\cos \theta_{\mu^+}$, $\cos \theta_{\mu^-}$, $\cos \theta_{\mu^+ \mu^-}$ in the inclusive analysis, after the preselection (2 well identified muons) and 120$<\mathrm{M}_{\mu^+ \mu^-}<$130 GeV requirements. All the distributions are normalized to 10. }

\end{center}

\subsection{Cut-count analysis}

\qquad The event numbers under selection flow are summarized in Table~\ref{tab:cc1}.
The two mass windows $\mathrm{M}_{\mu^+ \mu^-}$, $\mathrm{M}_{\text{recoil}}^{\mu^+\mu^-}$ are set in accord with the signal signature. The $\mathrm{P}_{\mathrm{T}_{\mu^+ \mu^-}}$, $\mathrm{P}_{\mathrm{Z}_{\mu^+ \mu^-}}$ are set to reduce the ZZ, where one of the Z boson decays to $\mu^+\mu^-$, and Drell-Yan Z$\to\mu^+\mu^-$ background.
The Higgs and Z boson decays can lead to different $\cos \theta _{\mu^+}$, $\cos \theta_{\mu^-}$ distributions due to the spin-dependence of the couplings and the parity violation of the Weak interaction. $\cos \theta _{\mu^+ \mu^-}$ selection is chosen to supress the 2f background.

\end{multicols}
  \begin{center}

  \tabcaption[Yield]{ Signal and background number of events under selection flow for the inclusive analysis. The simulation corresponds to CEPC at $\sqrt{s}=250$~GeV with an integrated luminosity of $5000\fbinv$. }
  \label{tab:cc1}
  \footnotesize
    \begin{tabular*}{130mm}{@{\extracolsep{\fill}}ccccccc}
      \toprule
      Category         &signal     &   ZZ    &    WW  & ZZorWW & SingleZ & 2f\\
      \hline
      Preselection   &     207.3 & 311312  & 129869 & 501590 & 63658  & 1740371\\
      120$<\mathrm{M}_{\mu^+ \mu^-}<$130   &     189.7 & 5479    & 17126  & 57405  & 1868   & 52525\\
      90.8$<$$\mathrm{M}_{\text{recoil}}^{\mu^+\mu^-}$$<$93.4 &  118.4 & 1207    & 868    & 2115   & 164   & 1157\\
      25$<\mathrm{P}_{\mathrm{T}_{\mu^+ \mu^-}}<$64      & 109.8 & 1009     & 725    & 1772   & 126    & 452 \\
      -56$<\mathrm{P}_{\mathrm{Z}_{\mu^+ \mu^-}}<$56   & 107.1 & 969     & 687    & 1726   & 120    & 420\\
      \tabincell{c}{$\cos\theta_{\mu^-}$$<$0.38 \\ $\cos\theta_{\mu^+}$$>$-0.38}        &  65.2   & 464     & 49     & 196    & 53     & 159\\
      $\cos\theta_{\mu^+ \mu^-}$$>$-0.996         &  65.0   & 462     & 46     & 196    & 52     & 99\\
      efficiency       &   31.3\%  &       &       &      &       & \\
      \bottomrule
    \end{tabular*}

  \end{center}

\begin{multicols}{2}

An unbinned maximum likelihood fit is performed on $\mathrm{M}_{\mu^+ \mu^-}$ distribution.
The signal is parameterized by a crystal ball function, with parameters fixed by simulated events.
The background is parametrized by a second order Chebychev function.

\begin{center}
  \includegraphics[width=6cm]{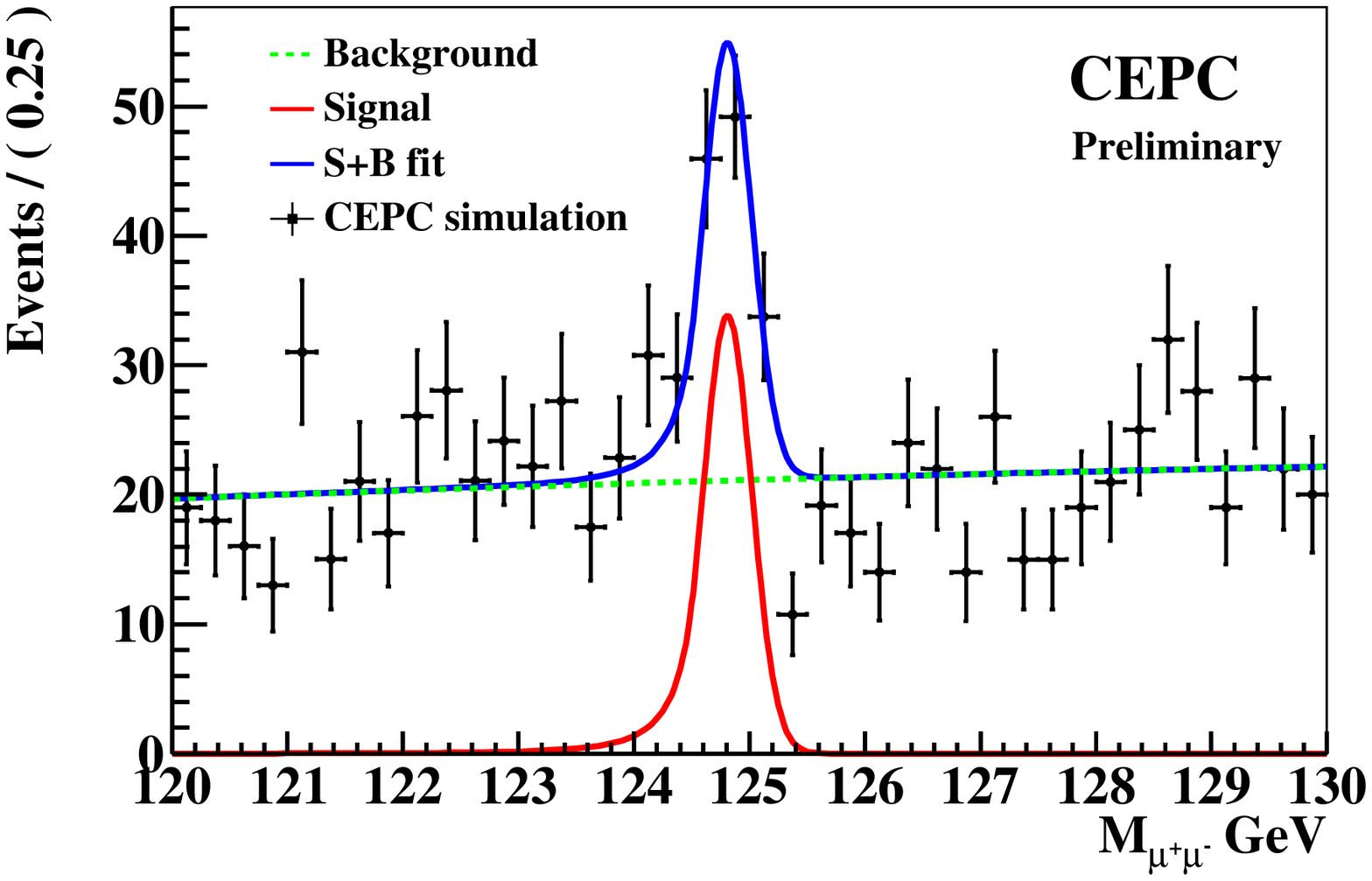}
  \figcaption{The invariant mass spectrum of di-muon system in the inclusive analysis . The dotted points with error bars represent data from CEPC simulation.The red-solid and green-dashed lines correspond to the signal and background contributions and
the solid-blue line represents the post-fit value of the total yield. }
  \label{fig:cc1}
\end{center}

Figure~\ref{fig:cc1} shows the post-fit result of the invariant mass distribution of the dimuon system. The fitted number of signal event is $77.2 \pm 13.0$. At 68\% confidence level, an accuracy from -17\% to 18\% on the signal strength can be achieved based on a likelihood scan.
The signal under the peak $124$-$125$~GeV leads to a high significance of 8.8 $\sigma$, via simple couting $\sqrt{2(s+b)ln(1+\frac{s}{b})-s}$, with $s$ and $b$ represent signal and background yields.

\subsection{BDT optimization}

\qquad
We have also exploited the Toolkit for Multivariate Analysis (TMVA)~\cite{TMVA} for further background rejection, where the  method of Gradient Boosted Decision Trees (BDTG) is adopted.  After fixing the range of the invariant mass and the recoil mass as mentioned above,  5 variables are taken as inputs to TMVA, including $\cos\theta_{\mu^{\pm}Z}$, $\cos\theta_{\mu^{\pm}}$ and $\mathrm{P}_{\mathrm{Z}_{\mu^+ \mu^-}}$. The choice of these variables are based on many tests and importance ranking. The resulted BDT response distribution can be seen in Figure~\ref{fig:mva1}, where the agreement between training and testing samples shows no obvious overtaining.  We then take the final event selections as:  BDTG response $>$ 0.369,  20$<$$\mathrm{P}_{\mathrm{T}_{\mu^+ \mu^-}}$$<$64~GeV and $\cos\theta_{\mu^+ \mu^-}>$ -0.996.  A maximum likelihood fit is performed on the resulted invariant mass of the di-muon system. The signal and background probability functions are parametrized in the same form as in the previous cut-count study.

\begin{center}
\begin{minipage}[c]{0.5\textwidth}
\centering
\includegraphics[angle=0,width=6cm]{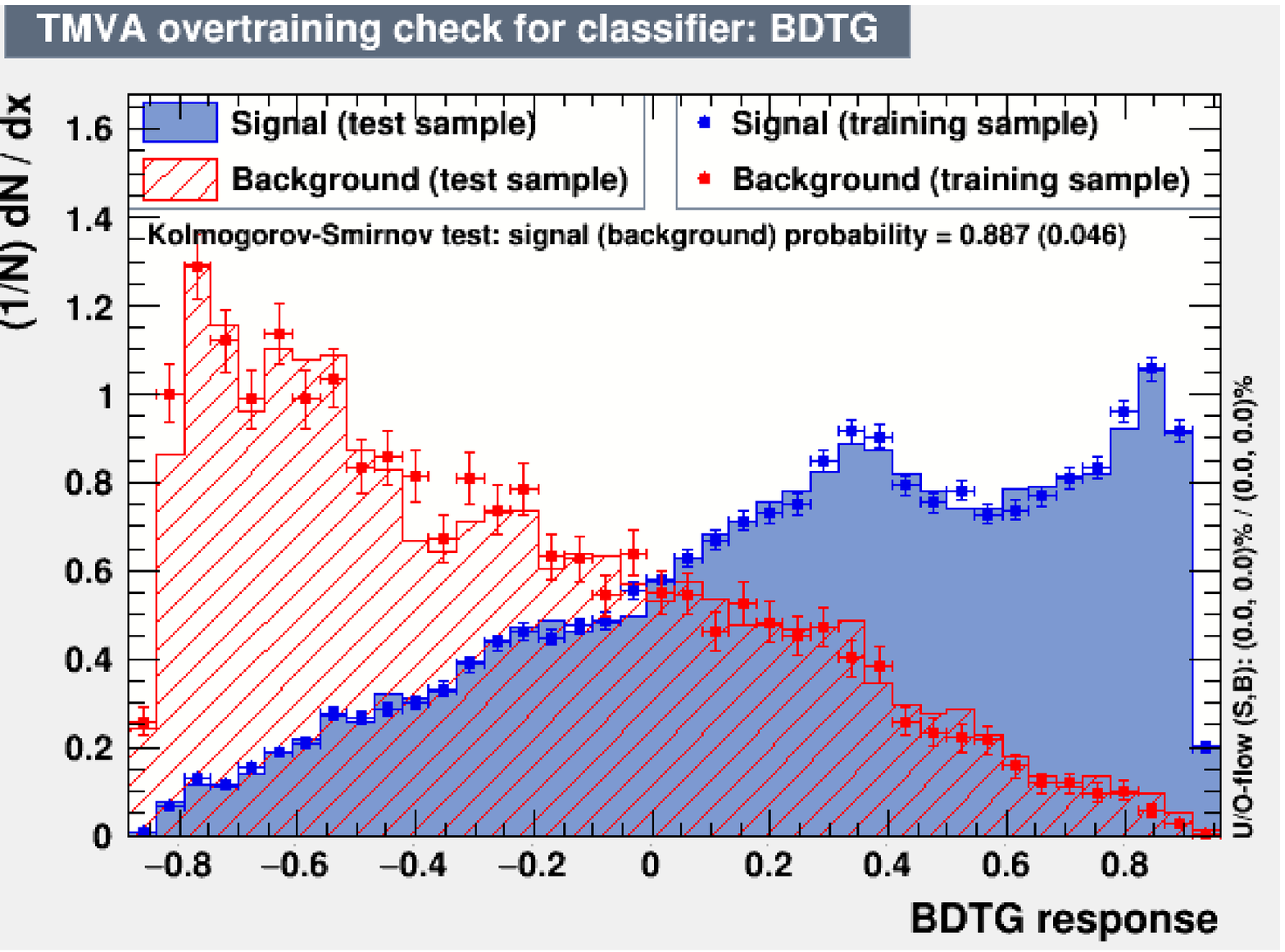}
\centering
\end{minipage}%

\begin{minipage}[c]{0.5\textwidth}
\centering
\includegraphics[angle=0,width=6cm]{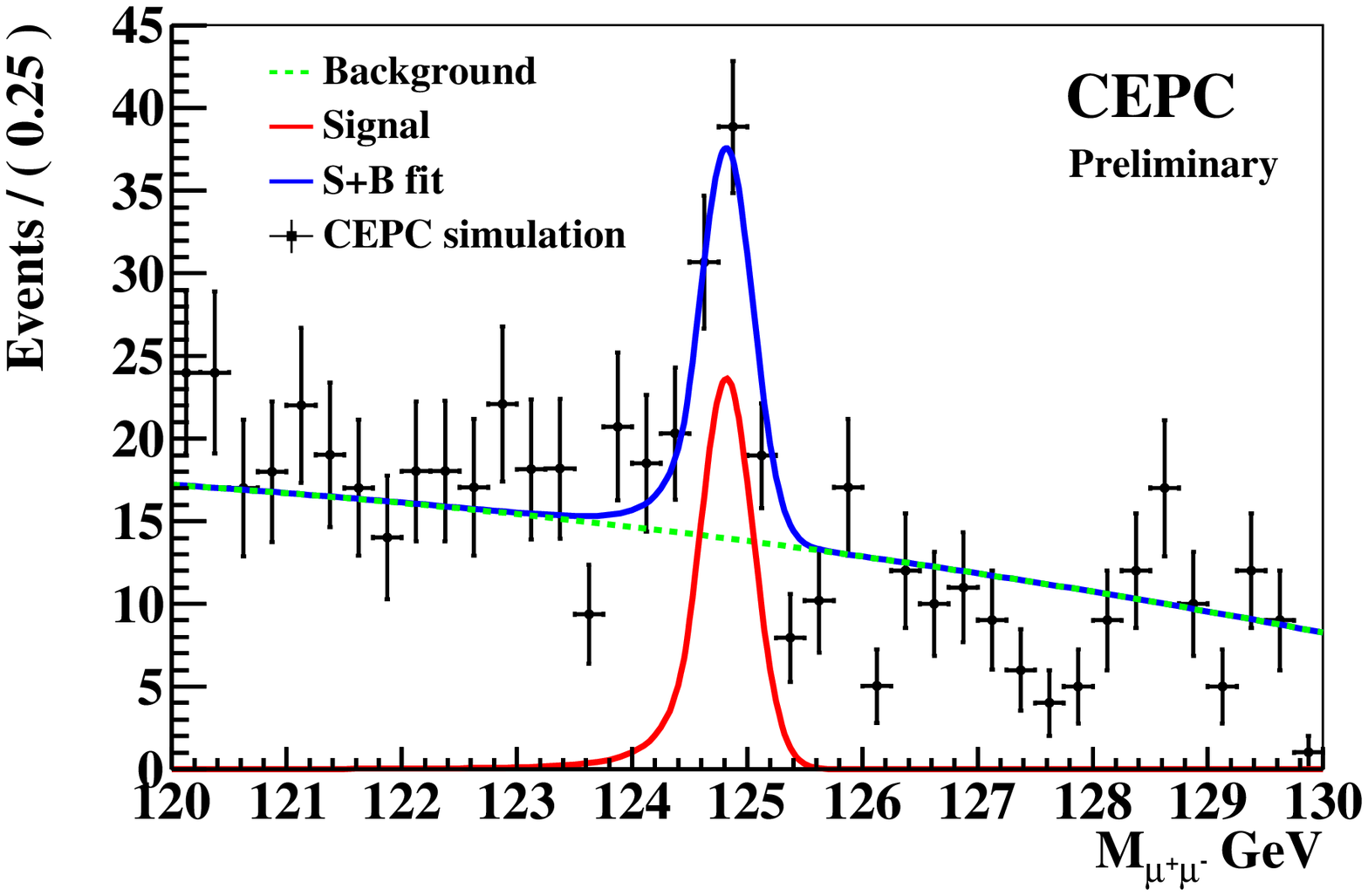}
\label{fig:mva1}
\end{minipage}

\figcaption{\label{fig:mva1}The BDT response distribution(top) and the post-fit result with BDT improvement(below). }

\end{center}

Figure~\ref{fig:mva1} shows the BDT response distribution and the post-fit result of $M_{\mu^+\mu^-}$.  The fitted number of the signal is $62.3 \pm 10.9$.
At 68\% confidence level, an accuracy from -16\% to 17\% on the signal strength can be achieved based on a likelihood scan. The signal under the peak $124$-$125$~GeV leads a significance of 10.9~$\sigma$.

\section{Z$(q\bar{q})\higgs(\mu\mu)$ analysis}
\label{exclusive}

\qquad Among all Z boson decay modes, hadronic channel is most promising due to its large branching fraction ($\sim 70$\%).  The exclusive method of kt algorithm for $e^+ e^-$ collisions in the Fastjet~\cite{Cacciari:2011ma} is used to reconstruct two jets with the particles expect the chosen $\mu^-$ and $\mu^+$, and the jets are sorted by energy. We perform an analysis on the Z$(q\bar{q})\higgs(\mu\mu)$ production. Apart from previously mentioned variables related to the $\higgs(\mu\mu)$ system, we further exploit the following selections on jets: third component of di-jet system momentum $\mathrm{P}_{\mathrm{Z}_{jj}}$, recoil mass of the di-jet system $\mathrm{M}_{\text{recoil}}^{jj}$
mass of jets $\mathrm{M}_{j1,2}$ and invariant mass of the di-jet system $\mathrm{M}_{jj}$.

\begin{center}
\begin{minipage}[c]{0.25\textwidth}

\includegraphics[width=4cm]{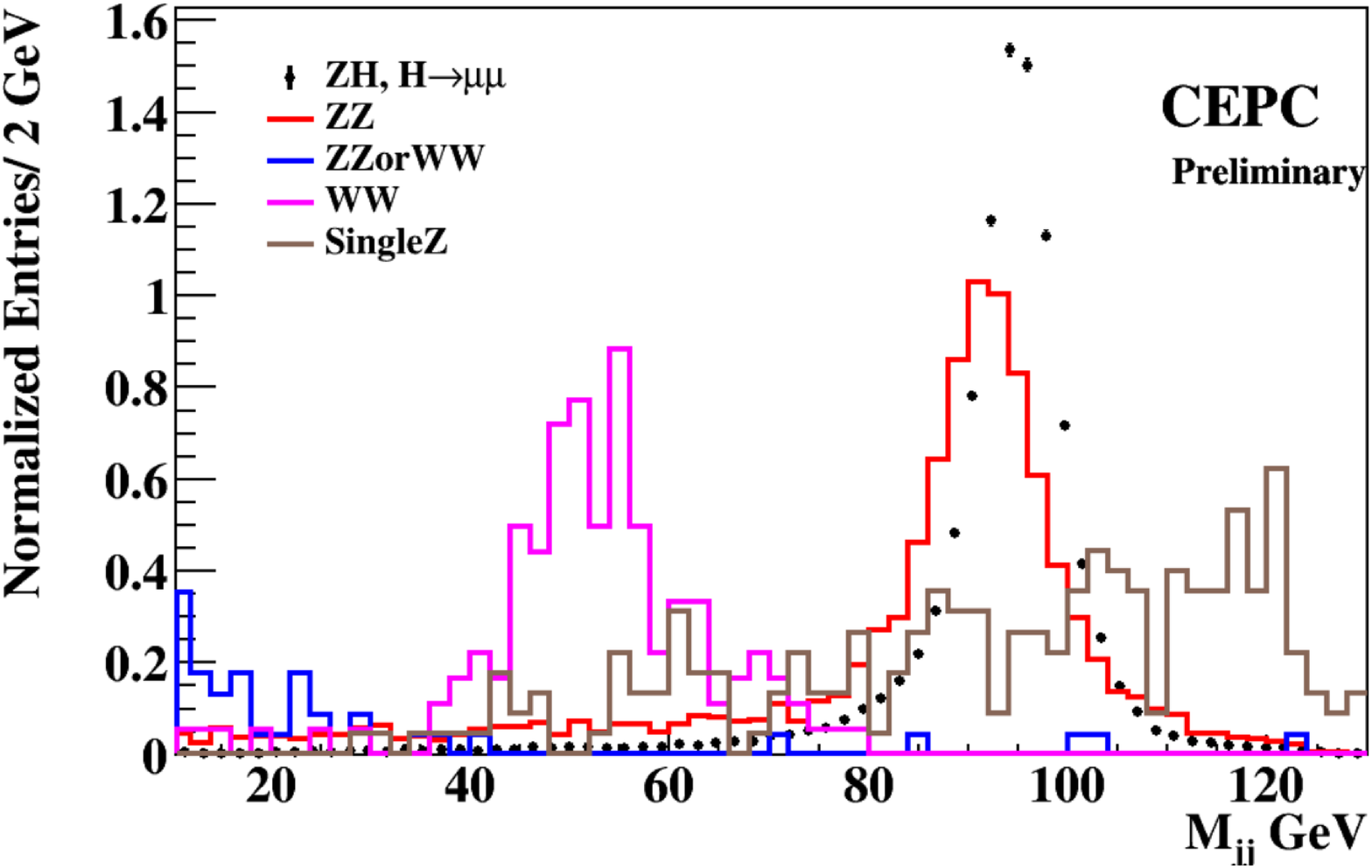}

\end{minipage}%
\begin{minipage}[c]{0.25\textwidth}

\includegraphics[width=4cm,]{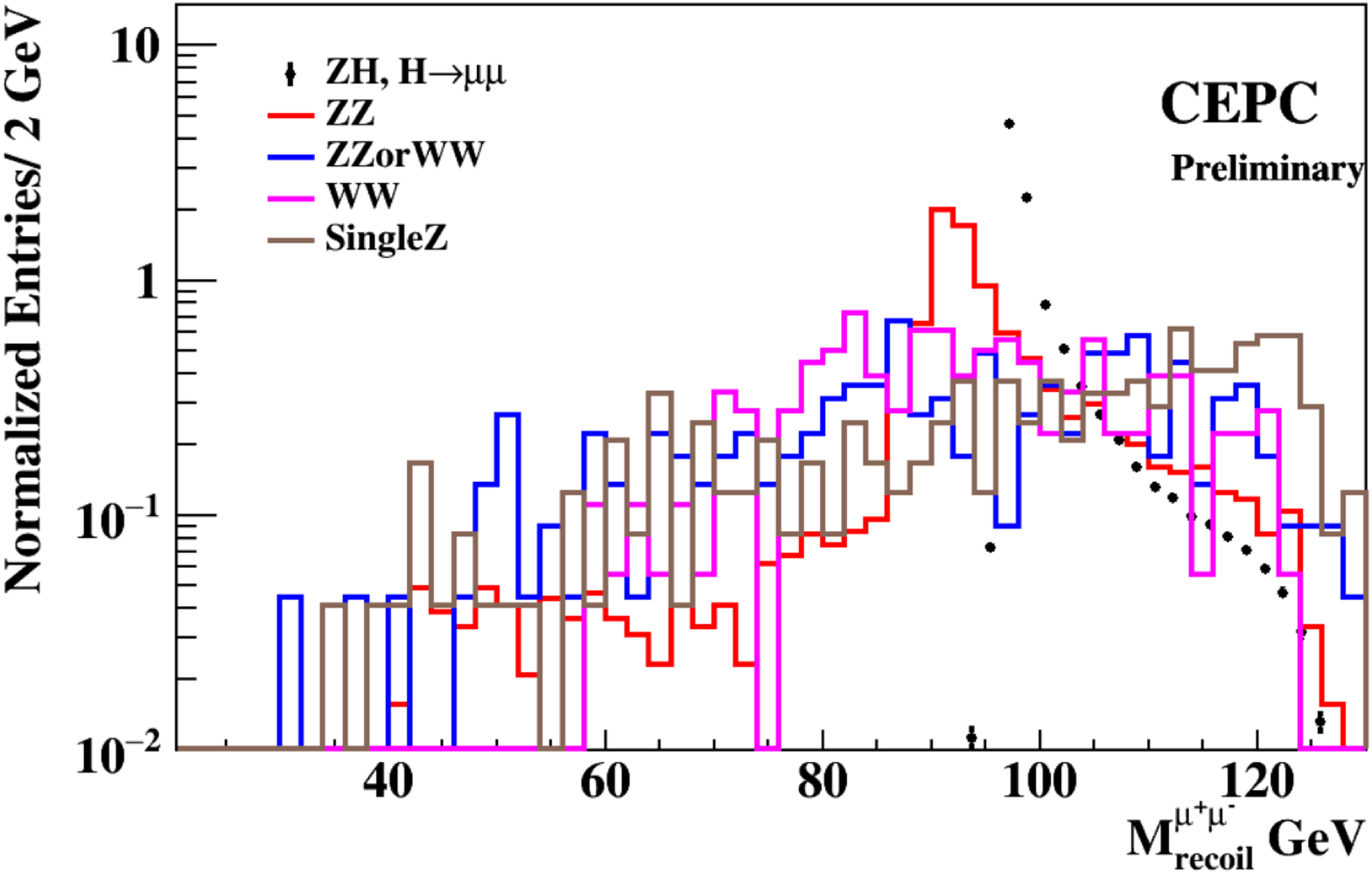}
\label{fig:cc2}
\end{minipage}
\figcaption{Distributions of the $\mathrm{M}_{\text{recoil}}^{\mu^+\mu^-}$, $\mathrm{M}_{jj}$ in Z$(q\bar{q})\higgs(\mu^+ \mu^-)$ analysis. And the distributions are normalize to 10. }
\end{center}

\subsection{Cut-count analysis}

\qquad A cut-count analysis is performed for the exclusive analysis. The event flow under selections are summarized in Table~\ref{tab:cc2}.  Selections on single and di-jet masses eliminates most background without hard jets. Recoil mass cut forther reduces the Z($ll$)Z(q$\bar{\mathrm{q}}$) background.

\end{multicols}

  \begin{center}
   \tabcaption{
    The cut-chain with cut-base method in the Z$(q\bar{q})\higgs(\mu\mu)$ analysis.}
  \label{tab:cc2}
  \footnotesize
    \begin{tabular*}{130mm}{@{\extracolsep{\fill}}ccccccc}

      \toprule
      Category                      &signal     &   ZZ    &    WW  & ZZorWW & SingleZ & 2f\\
      \hline
      Preselection                  &     156.3 & 390775  & 183751 & 463361 & 101164  & 0\\
      120$<\mathrm{M}_{\mu^+ \mu^-}<$130    &     141.6 & 3786    & 181    & 227    & 244    & 0\\
      \tabincell{c}{$\mathrm{M}_{j1}$$<$4.2 \\ $\mathrm{M}_{j2}$$<$2.8} &  133.0  & 3216   & 111    & 0 & 9  & 0\\
      $\mathrm{M}_{jj}$$>$76.0                   &     127.5 & 2917    & 2    & 0     & 8       & 0\\
      90.9$<$$\mathrm{M}_{\text{recoil}}^{\mu^+\mu^-}$$<$93.5         &     75.2  & 893     & 0    & 0    & 0      & 0\\
      20$<$$\mathrm{P}_{\mathrm{T}_{\mu^+ \mu^-}}$$<$64             &     74.5  & 777     & 0    & 0    & 0      & 0\\
      -58$<$$\mathrm{P}_{\mathrm{Z}_{\mu^+ \mu^-}}$$<$58              &     74.5  & 748     & 0    & 0    & 0      & 0\\
      \tabincell{c}{$\cos\theta_{\mu^+}$$>$-0.98 \\  $\cos\theta_{\mu^-}$$<$0.98 }       &     74.2  & 747     & 0    & 0    & 0      & 0\\
      efficiency       &   47.5\%  &       &       &      &       & \\
      \bottomrule
    \end{tabular*}

  \end{center}

\begin{multicols}{2}

As in the inclusive channel, we perform a likelihood fit to extract the signal yield and strength parameter. Quality of the fit is demonstrate in Fig.~\ref{fig:fit2cc}.
The signal yield from the fit is $75.5 \pm 12.5$.
The signal strength can be determined with an uncertainty from -16\% to 17\%, at 68\% confidence level. The signal significance under the peak 124-125~GeV is found to be 10.8$\sigma$.

\begin{center}
  \includegraphics[width=6cm]{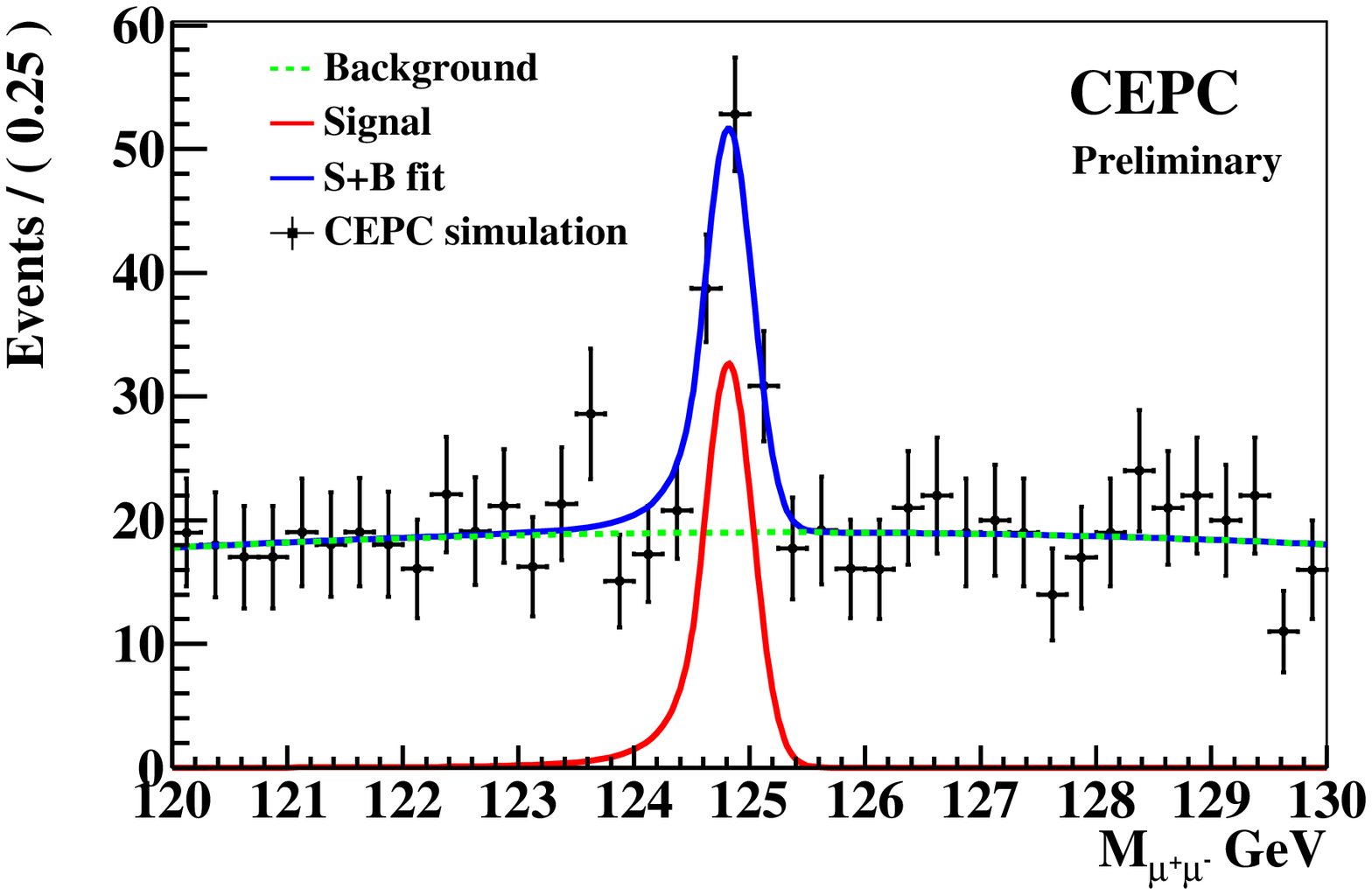}
  \figcaption{The invariant mass spectrum of di-muon system in the $Z(qq)H( \mu \mu)$ analysis. The dotted points with error bars represent data from CEPC simulation. The red-solid and green-dashed lines correspond to the signal and background contributions and the solid-blue line represents the post-fit value of the total yield.}
  \label{fig:fit2cc}
\end{center}

\subsection{BDT improvement}
\qquad In order to achieve highest significance, we perform a two step multivariate analysis.
The first step exploit a MLP (Multilayer Perceptron)[26] method to suppress the fully leptonic WW and ZZ backgrounds. After applying $\mathrm{M}_{\text{recoil}}^{\mu^+\mu^-}$$>$ 90 GeV, 4 variables including $\mathrm{M}_{j1,2}$,  $\mathrm{M}_{jj}$ and $\mathrm{M}_{\text{recoil}}^{jj}$   are considered as inputs for the MLP. The effectiveness of this MLP is shown in Fig.~\ref{fig:mlp}.
After requiring MLP response to be greater than 0.71, we exploit BDTG to further reduce the backgounds from semileptonic ZZ and WW. In this second step, variables $\cos\theta_{\mu^{\pm}}$, $\cos\theta_{\mu^{\pm}\mathrm{Z}}$,  $\mathrm{P}_{\mathrm{Z}_{\mu^+ \mu^-}}$, $\mathrm{P}_{\mathrm{Z}_{jet12}}$, $\cos\theta_{j1/j2,H}$, $\cos\theta_{j1,2}$, $\mathrm{M}_{jj}$ are taken as inputs.

\begin{center}

  \includegraphics[width=6cm]{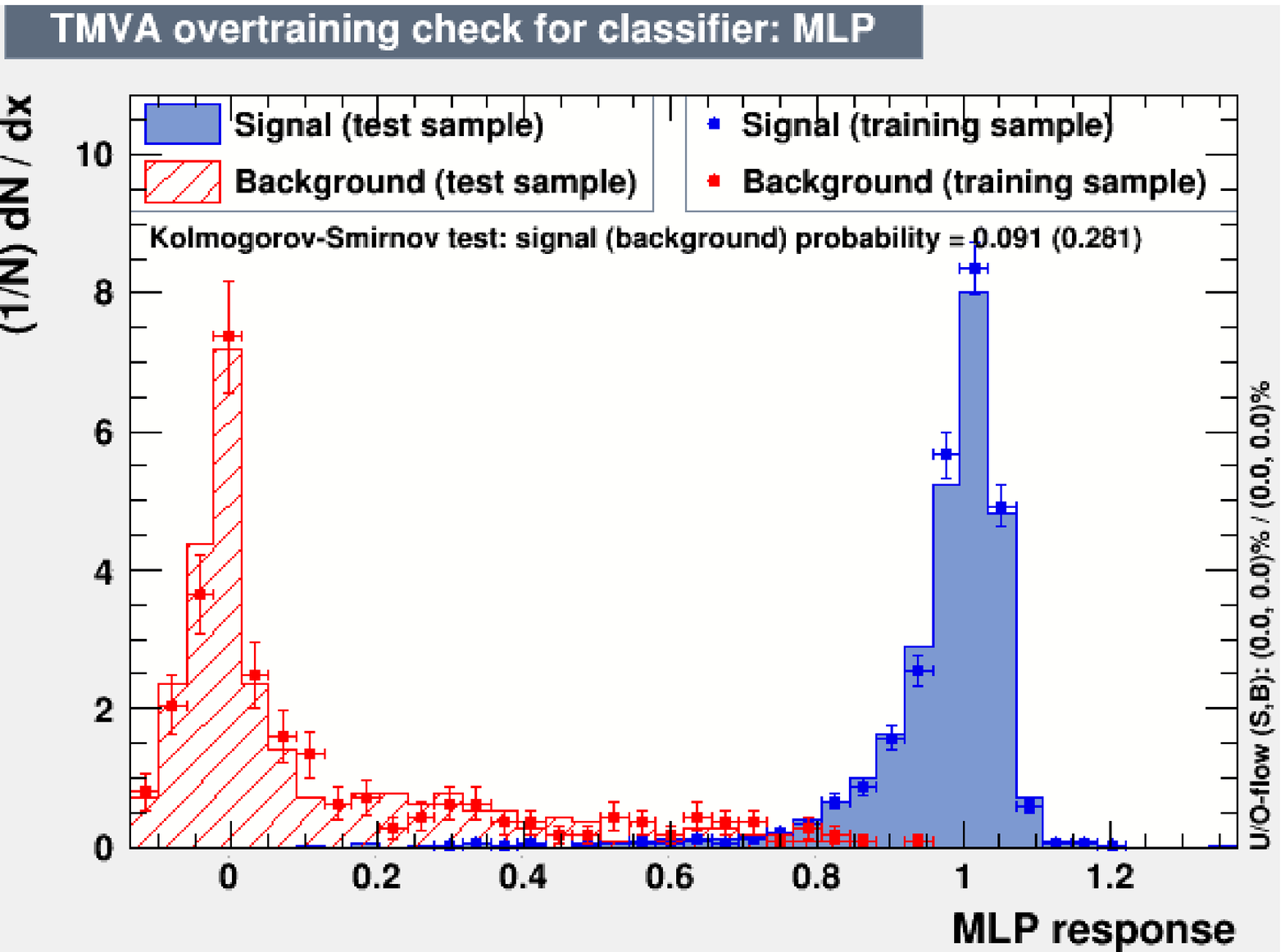}

  \figcaption{The MLP result and the overtraining test in the $Z(qq)H( \mu \mu)$ analysis. }
  \label{fig:mlp}

\end{center}

\begin{center}
\begin{minipage}[c]{0.5\textwidth}
\centering
\includegraphics[angle=0,width=6cm]{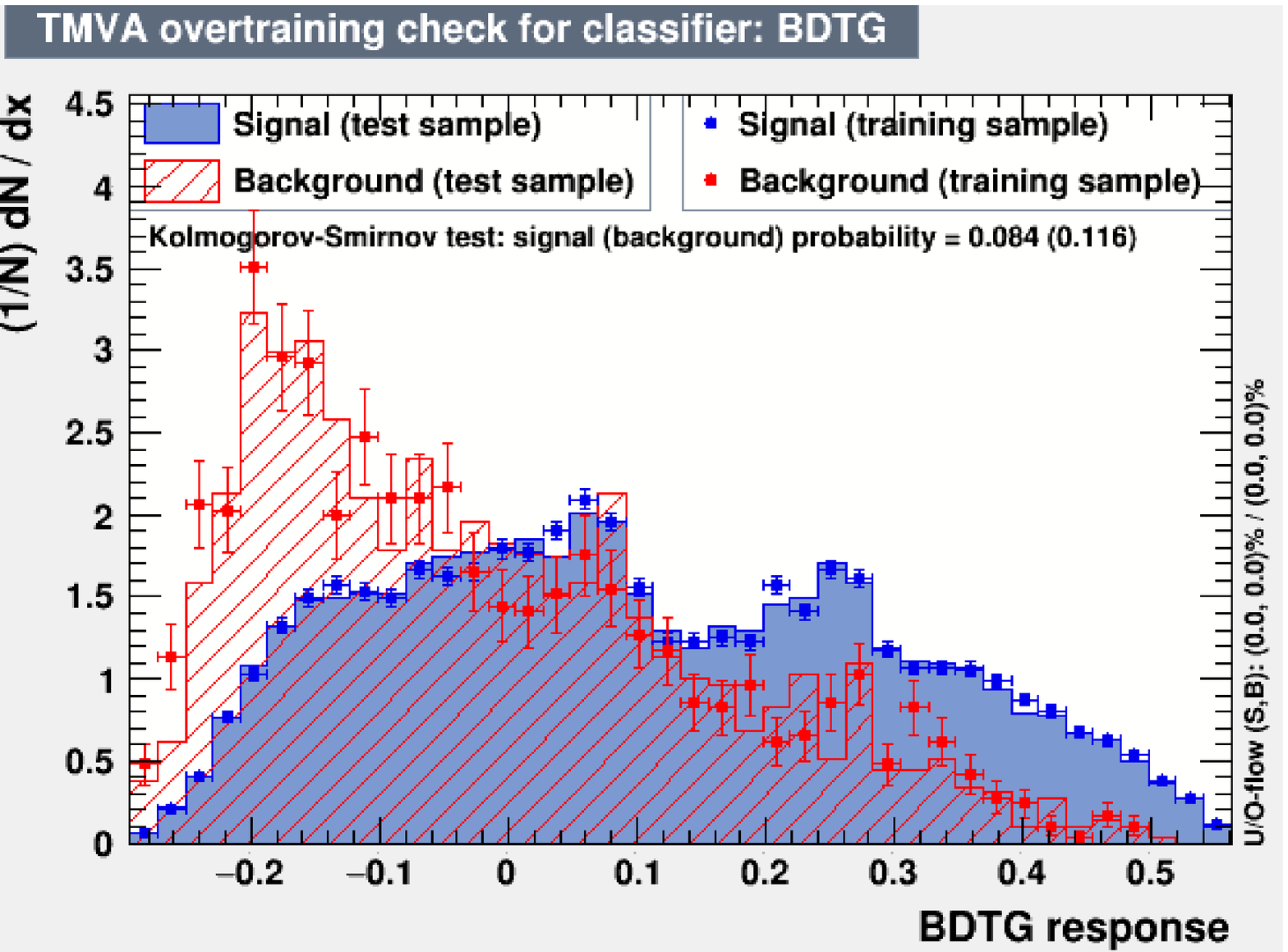}
\centering
\end{minipage}%

\begin{minipage}[c]{0.5\textwidth}  \centering\includegraphics[angle=0,width=6cm]{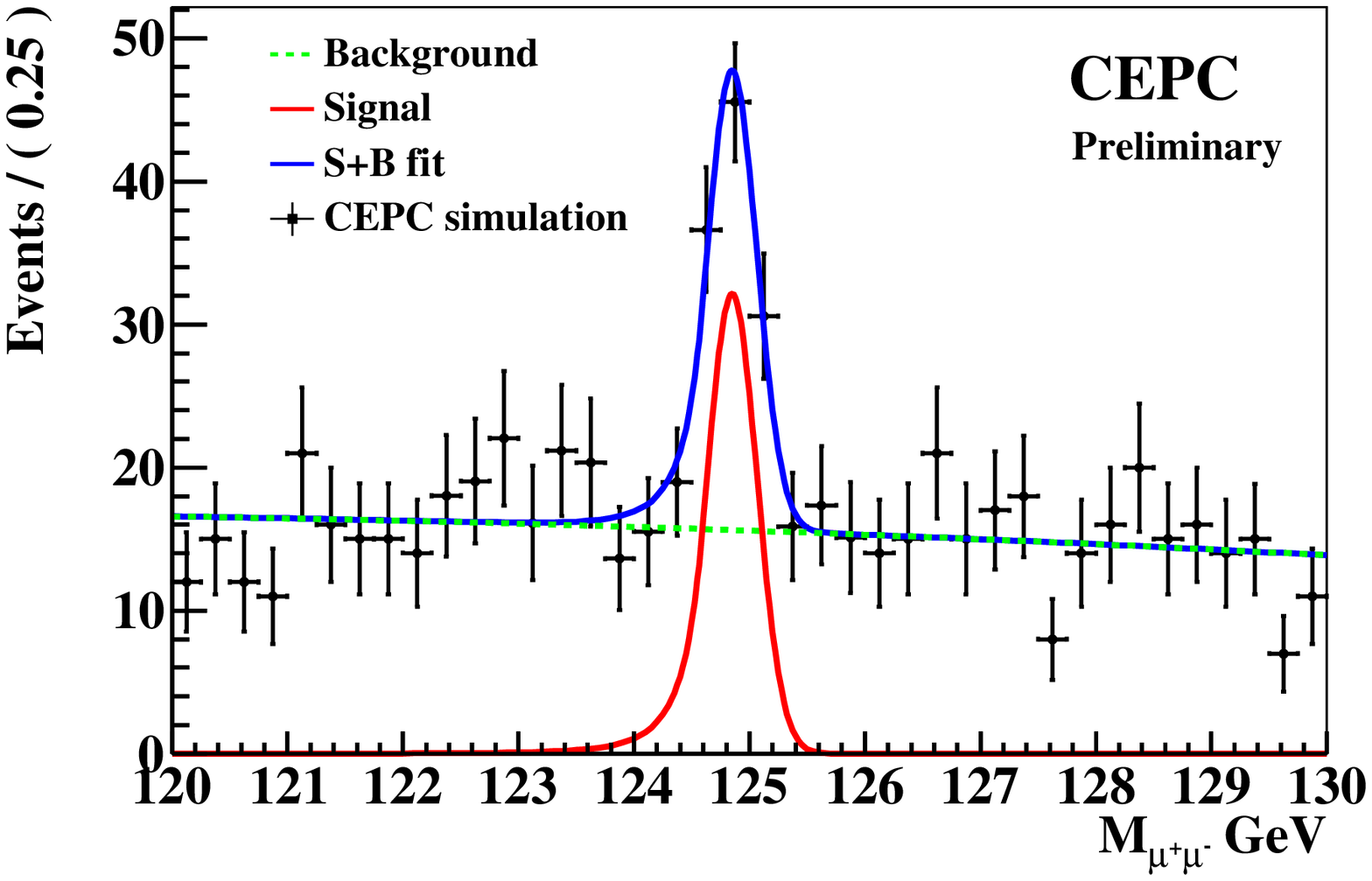}
\end{minipage}
\end{center}
\figcaption{The BDT response(top) and the final fit result(below) in the Z$(qq)H( \mu \mu)$ channel analysis }
\label{fig:mva2}

After the two step multivariate analysis, we require BDTG response$>$-0.13, 90.4 $<$ $\mathrm{M}_{\text{recoil}}^{\mu^+\mu^-}$ $<$ 93 GeV and 28$<$$P_{\mathrm{T}_{\mu^+ \mu^-}}$$<$64 GeV.  Finally, we perform a likelihood fit to extract the signal yield and
strength parameter, as shown in Fig.~\ref{fig:mva2}.  The signal yield from the fit is $73.4 \pm 12.4$.
Based on a likelihood scan, the signal strength can be determined with an uncertainty from -16\% to 17\%, at 68\% confidence level.  The significance of the signal in the peak region 124-125~GeV is found to be 10.8$\sigma$.

\section{Summary}
\label{summary}

\qquad Feasibility of measuring $\higgs \to \mathrm{\mu^+ \mu^-}$ at the CEPC is studied
considering a center of mass energy $250$ GeV collision and 5000 fb$^{\rm{-1}}$ integrated liminosity.
The measurement is perfomed in two complementary channels: ZH production without measuring the Z boson decay and ZH production with the Z boson hadronically decay.
For each decay channel, a cut-count analysis is tested and followed with an improvement using multivariate techniques.
Similar results are obtained from two channels.
Over 10~$\sigma$ significance can be reached for the signal $\higgs \to \mathrm{\mu^+ \mu^-}$
process.
Accuracy of the signal strength can be measured with $\pm$14\% uncertainty and
the associated H-$\mathrm{\mu}$-$\mathrm{\mu}$ coupling can be restricted to 10\% level.
The results are comparable to the High-Luminosity LHC.

~\\

\acknowledgments{The authors would like to thank Xin Mo, Dan Yu and Yuqian Wei for useful discussions. This work is supported in part by the National Natural Science Foundation of China, under Grants No. 11475190 and No. 11575005,  by the CAS Center for Excellence in Particle Physics (CCEPP), and by CAS Hundred Talent Program (Y3515540U1)}

\end{multicols}

\vspace{10mm}

\begin{multicols}{2}

\end{multicols}

\vspace{-1mm}
\centerline{\rule{80mm}{0.1pt}}
\vspace{2mm}

\subsection*{Appendices A}

  \begin{center}
  \tabcaption[Monte Carlo purities in the single lepton sample]{The information of the two fermions background samples}
    \begin{tabular}{ccccccc}
      \toprule
     Process  & Final states & $\sigma$ $[\mathrm{fb}]$ & Events expected \\ \hline
        uu & $u, \bar{u}$ & 9995.35 & 50476527  \\
        dd & $d, \bar{d}$ & 9808.71 & 49533965  \\
        cc & $c, \bar{c}$ & 9974.20 & 50369725  \\
        ss & $s, \bar{s}$ & 9805.39 & 49517234  \\
        bb & $b, \bar{b}$ & 9803.04 & 49505372 \\ \hline
        qq & $q, \bar{q}$ & 49561.30 & 250284565  \\ \hline
        e2e2 & $\mu^-\mu^+$ & 4967.58 & 25086253  \\
        e3e3 & $\tau^-\tau^+$ & 4374.94 & 22093447  \\
        bhabha & $e^-, e^+, \gamma$ & 24992.21  & 126210660  \\
      \bottomrule
    \end{tabular}

  \label{tab:example1}
  \end{center}

~\\
~\\
~\\
~\\
~\\
~\\
~\\
~\\
~\\
~\\
~\\
~\\
~\\
~\\
~\\

  \begin{center}
  \tabcaption[Monte Carlo purities in the single lepton sample]{The information of the four fermions background samples}
    \begin{tabular}{ccccccc}
      \toprule
      Process        &	 Final states&	 $\sigma$ $[\mathrm{fb}]$&	 Events expected\\	\hline
ZZ(h)utut&	 $up, up, up, up$&	83.09&	419604\\	
ZZ(h)dtdt&	 $down, down, down, down$&	226.2&	1142310\\	
ZZ(h)uu\_notd&	 $uq, uq, (sq, bq), (sq, bq)$&	95.65&	483032\\	
ZZ(h)cc\_nots&	 $cq, cq, (dq, bq), (dq, bq)$&	96.04&	485002\\	\hline
ZZ(sl)nu\_up&	 $nu_{\mu,\tau}, nu_{\mu,\tau}, up, up$&	81.72&	412686\\	
ZZ(sl)nu\_down&	 $nu_{\mu,\tau}, nu_{\mu,\tau}, down, down$&	134.86&	681043\\	
ZZ(sl)mu\_up&	 $mu, mu, up, up$&	82.38&	416019\\	
ZZ(sl)mu\_down&	 $mu, mu, down, down$&	127.96&	646198\\	
ZZ(sl)tau\_up&	 $tau, tau, up, up$&	39.78&	200889\\	
ZZ(sl)tau\_down&	 $tau, tau, down, down$&	64.3&	324715\\	\hline
ZZ(l)4tau&	 $\tau^-, \tau^+, \tau^-, \tau^+$&	4.38&	22119\\	
ZZ(l)4mu&	 $\mu^-, \mu^+, \mu^-, \mu^+$&	14.57&	73578\\	
ZZ(l)taumu&	 $\tau^-, \tau^+, \mu^-, \mu^+$&	17.54&	88577\\	
ZZ(l)mumu&	 $\nu_{\tau}, \bar{\nu}_{\tau}, \mu^-, \mu^+$&	18.17&	91758\\	
ZZ(l)tautau&	 $\nu_{\mu}, \bar{\nu}_{\mu}, \tau^-, \tau^+$&	9.2&	46460\\	\hline
WW(h)cuxx&	 $uq, cq, down, down$&	3395.48&	17147189\\	
WW(h)uubd&	 $uq, uq, dq, bq$&	0.05&	252\\	
WW(h)uusb&	 $uq, uq, sq, bq$&	165.94&	837997\\	
WW(h)ccbs&	 $cq, cq, sq, bq$&	5.74&	28987\\	
WW(h)ccds&	 $cq, cq, sq, dq$&	165.57&	836128\\	\hline
WW(sl)muq&	 $mu, nu, up, down$&	2358.69&	11911394\\	
WW(sl)tauq&	 $tau, nu, up, down$&	2351.98&	11877519\\	
WW(l)ll&	 $mu, tau, nu_{\mu}, nu_{\tau}$&	392.96&	1984448\\	\hline
ZZorWW(h)udud&	 $uq, uq, dq, dq$&	1570.4&	7930514\\	
ZZorWW(h)0cscs&	 $cq, cq, sq, sq$&	1568.94&	7923141\\	
ZZorWW(l)mumu&	 $mu, mu, nu_{\mu}, nu_{\mu}$&	214.81&	1084790\\	
ZZorWW(l)tautau&	 $tau, tau, nu_{\tau}, nu_{\tau}$&	205.84&	1039492\\	\hline
sZ(l)etau&	 $e^-, e^+, \tau^-, \tau^+$&	150.14&	758207\\	
sZ(l)emu&	 $e^-, e^+, \mu^-, \mu^+$&	852.18&	4303527\\	
sZ(l)enu&	 $e^-, e^+, \nu_{\mu,\tau}, \bar{\nu}_{\mu,\tau}$&	29.62&	149581\\	
sZ(sl)eut&	 $e, e, up, up$&	195.86&	989093\\	
sZ(sl)edt&	 $e, e, down, down$&	128.72&	650036\\	\hline
sZ(l)numu&	 $\nu_e, \bar{\nu}_e, \mu^-, \mu^+$&	43.33&	218816\\	
sZ(l)nutau&	 $\nu_e, \bar{\nu}_e, \tau^-, \tau^+$&	14.57&	73578\\	
sZ(sl)nu\_up&	 $\nu_e, \bar{\nu}_e, up, up$&	56.09&	283254\\	
sZ(sl)nu\_down&	 $\nu_e, \bar{\nu}_e, down, down$&	91.28&	460964\\	\hline
sW(l)mu&	 $e, nu_e, mu, nu_{\mu,\tau}$&	429.2&	2167446\\	
sW(l)tau&	 $e, nu_e, tau, nu_{\mu,\tau}$&	429.42&	2168556\\	
sW(sl)qq&	 $e, nu_e, up, down$&	2579.31&	13025535\\	\hline
sWorsW(l)el&	 $e^-, e^+, \nu_e, \bar{\nu}_e$&	249.34&	1259167\\	

\bottomrule
    \end{tabular}

  \label{tab:example2}
  \end{center}

\clearpage

\end{CJK*}
\end{document}